\documentstyle[twocolumn,aps,psfig,amssymb]{revtex}
\begin{document}
\title{Comment on ``Using Fermi Statistics to Create Strongly Coupled Ion
Plasmas in Atom Traps''}
\author{M.~Bonitz, and D.~Semkat}
\address{Fachbereich Physik, Universit{\"a}t Rostock,
Universit{\"a}tsplatz 3, D-18051 Rostock, Germany}

\date{\today}
\maketitle
\begin{abstract}
The fermionic exchange energy of ultracold ions is computed. It is
shown that this effect allows to increase the ion coupling in traps
by about an order of magnitude compared to the classical case.
\end{abstract}

\vspace{2cm}
In this Comment we analyze the prospects of producing a strongly coupled
ion plasma by laser ionization of
an ultracold gas. A recent Letter \cite{murillo01} suggested that the Fermionic
exchange of the ensemble of atoms (before ionization) would allow to increase the
ionic coupling drastically. While this general idea is correct, we show that
the predicted ``orders of magnitude'' effect may only appear in a transient
regime. For the finite state coupling an upper bound is derived which is much
lower.

Ref. \cite{murillo01} assumes that after ionization the electrons equilibrate
instantaneously whereas the ions form a one-component plasma
with a statically screened Coulomb interaction (screening
parameter $\kappa=1/r_D=\sqrt{4\pi n_e e^2/kT_e}$ determined by the electrons).
Initially ion equilibration proceeds under
isolated conditions, i.e. total energy $E$ remains constant
\begin{eqnarray}
E(t) \equiv K(t)+U_F(t)+U(t) = K_0+U_{F0}+U_0 \equiv E_0.
\label{econs}
\end{eqnarray}
Here, $K$, $U_F$ and $U$ are, respectively, the kinetic, exchange (mean field or Fock) and
correlation energy of the ions, and ``$0$'' denotes the initial state.

If the initial state is uncorrelated, $U_{F0}=U_0=0$, the system
is heating up \cite{bonitz-etal.96jpcm}, $K(t)\ge K_0$, as a consequence
of growing attractive (negative) correlation energy, $U(t)<0$.
Ref. \cite{murillo01} suggested to use ionic exchange, $U_{F0}<0$
to reduce the heating. This is reasonable since atom cooling below the Fermi temperature
has already been achieved experimentally \cite{demarco99}.

\begin{figure}[p]
\centerline{
\psfig{file=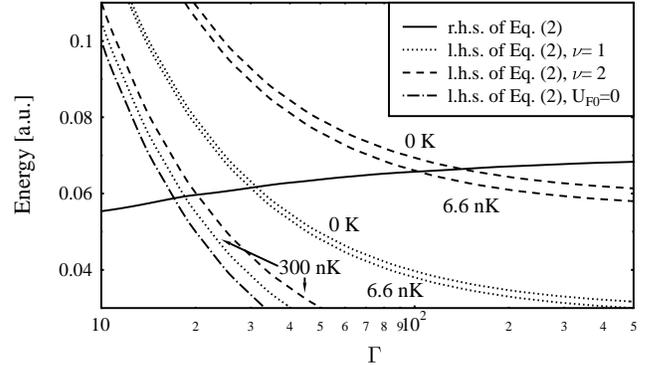,height=4.8cm}
}
\vspace{0.2cm}
\caption[]{\label{f1}
Graphical solution of Eq.~(\ref{gammaf}). The final state $\Gamma$ is obtained as
the crossing of the full line [r.h.s. of Eq.~(\ref{gammaf})] and one of the decreasing
curves [l.h.s. of Eq.~(\ref{gammaf})] for varying initial temperature and spin
polarization.
}
\end{figure}

For a {\em homogeneous} plasma, the {\em final state} temperature $T$ or coupling $\Gamma\equiv U_c/k_BT$
($U_c=e^2/a$ is the mean Coulomb energy and $a$ the mean interparticle distance),
can be estimated from Eq.~(\ref{econs}) by setting $U_0\rightarrow 0$,
\begin{eqnarray}
\frac{1}{\Gamma}+\frac{|U_{F0}(T_0)|-|U_{F}(T)|-k_BT_0}{U_c} =
\frac{2}{3}\left(\frac{|u|}{\Gamma}-\frac{\kappa}{2}\right),
\label{gammaf}
\end{eqnarray}
where the r.h.s. is the final state correlation energy $U$ (in units of $U_c$), cf.
\cite{murillo01,hamaguchi}.
Solutions of Eq.~(\ref{gammaf}) are shown in Fig.~\ref{f1} for potassium ions with density
$n=1.1\times 10^{11}$cm$^{-3}$ and $\kappa=4.5/a$ \cite{murillo01}. Note that the Fock energy $U_{F0}$
differs from the familiar exchange energy as it involves the screened interaction instead of
the Coulomb potential and is computed numerically. For the initial temperature $T_0$ we choose
$300$nK (corresponding to $T=0.5T_F$ of ions in a trap \cite{demarco99}) and $6.6$nK ($T=0.5T_F$
in a homogeneous system studied in \cite{murillo01}).
The highest possible effect can be estimated from the limit $T_0\rightarrow 0$ for which the
exchange energy can be found analytically,
\begin{eqnarray}
\frac{U^{\nu}_F(0,\kappa)}{\nu \cdot U_c}=\frac{1}{4}\left(\frac{3}{2\pi}\right)^{2/3}
\bigg\{
3-2\alpha^2-8\alpha \,\arctan{\frac{1}{\alpha}}
\nonumber\\
+2\alpha^2(3+\alpha^2)\ln{\left(1+\frac{1}{\alpha^2}\right)}
\bigg\}, \quad \mbox{with} \quad \alpha=\frac{\kappa}{2 k_F}   \quad,
\end{eqnarray}
where $k_F$ is the Fermi wave number and $\nu=2(1)$ for spin polarized (un-polarized) ions.

Our analysis reveals that the final state $\Gamma$ indeed increases due to exchange effects,
from $17$ (without correlations) up to $\Gamma \simeq 140$, for $T_0=0$ and spin polarized ions,
cf. Fig.~\ref{f1}. To achieve this, in homogeneous systems, requires temperatures below
$10$nK whereas, in traps, several $100$nK should be sufficient.


This work is supported by the Deutsche Forschungsgemeinschaft
(grant BO-1366/3) and the NIC J\"ulich.

\end{document}